\documentstyle[12pt]{article}
\begin{document}

\def\bfr{\begin{flushright}}
\def\efr{\end{flushright}}
\def\bfl{\begin{flushleft}}
\def\efl{\end{flushleft}}
\def\si{\quad}
\def\sii{\qquad}
\def\siii{\qquad \qquad}
\def\noi{\noindent}
\def\vs{\vspace}
\def\vss{\vspace{.5cm}}
\def\vsss{\vspace{1cm}}
\def\vssss{\vspace{1.5cm}}
\def\begc{\begin{center}}
\def\endc{\end{center}}
\def\bmp{\begin{minipage}}
\def\emp{\end{minipage}}
\def\sct#1{\setcounter{equation}{#1}}
\def\bena{\begin{eqnarray}}
\def\benan{\begin{eqnarray*}}
\def\eena{\end{eqnarray}}
\def\eenan{\end{eqnarray*}}
\def\nn{\nonumber  \\}
\def\sct#1{\setcounter{equation}{#1}}
\def\qed{\hfill \hbox{\rule{6pt}{6pt}}}
\def\rar{\rightarrow}
\def\lar{\leftarrow}
\def\begs{\begin{screen}}
\def\ends{\end{screen}}
\def\rarw{\rightarrow}
\def\rarww{\longrightarrow}
\def\larw{\leftarrow}
\def\larww{\longleftarrow}
\def\larww{\longleftarrow}
\def\lrarw{\leftrightarrow}
\def\lrarww{\longleftrightarrow}
\def\mb{\makebox[3cm]}
\def\mbc{\makebox[2cm][c]}
\def\mbq{\makebox[1.05cm][l]}
\def\mbr{\makebox[2cm][r]}
\def\mbl{\makebox[2cm][l]}
\def\mBc{\makebox[2.5cm][c]}
\def\mBr{\makebox[2.5cm][r]}
\def\mBl{\makebox[2.5cm][l]}
\def\mbbc{\makebox[4cm][c]}
\def\mbbr{\makebox[4cm][r]}
\def\mbbl{\makebox[4cm][l]}
\def\mbbbc{\makebox[6cm][c]}
\def\mbbbr{\makebox[6cm][r]}
\def\mbbbl{\makebox[6cm][l]}
\def\mbbbbc{\makebox[8cm][c]}
\def\mbbbbr{\makebox[8cm][r]}
\def\mbbbbl{\makebox[8cm][l]}
\def\Mfr{\makebox[12cm][c]{} \mbl}
\def\und{\underline}
\def\a{\alpha}
\def\b{\beta}
\def\g{\gamma}
\def\G{\Gamma}
\def\d{\delta}
\def\D{\Delta}
\def\e{\epsilon}
\def\ve{\varepsilon}
\def\z{\zeta}
\def\h{\eta}
\def\th{\theta}
\def\Th{\Theta}
\def\vt{\vartheta}
\def\i{\iota}
\def\k{\kappa}
\def\l{\lambda}
\def\L{\Lambda}
\def\m{\mu}
\def\n{\nu}
\def\x{\xi}
\def\X{\Xi}
\def\o{\o}
\def\p{\pi}
\def\P{\Pi}
\def\vp{\varpi}
\def\r{\rho}
\def\vr{\varrho}
\def\s{\sigma}
\def\Sg{\Sigma}
\def\t{\tau}
\def\u{\upsolon}
\def\U{\Upsilon}
\def\f{\phi}
\def\F{\Phi}
\def\vf{\varphi}
\def\c{\chi}
\def\ps{\psi}
\def\Ps{\Psi}
\def\w{\omega}
\def\W{\Omega}
\def\bra{\langle}
\def\ket{\rangle}
\def\fr{\frac}
\def\del{\partial}
\def\deldel{\del _{\m}\del ^{\m}}
\def\dt{\fr{\del}{\del t}}
\def\dx{\fr{\del}{\del x}}
\def\dxo{\fr{\del}{\del x^0}}
\def\dxi{\fr{\del}{\del x^1}}
\def\dxii{\fr{\del}{\del x^2}}
\def\dxiii{\fr{\del}{\del x^3}}
\def\ddx{\fr{\del ^2}{\del x^2}}
\def\Sr{Schr\"{o}dinger}
\def\mxxb{\left( \begin{array}{cc}}           
\def\mxxe{\end{array} \right)}
\def\mxxxb{\left( \begin{array}{ccc}}         
\def\mxxxe{\end{array} \right)}
\def\mxxxxb{\left( \begin{array}{cccc}}       
\def\mxxxxe{\end{array} \right)}
\def\mxxxxxb{\left( \begin{array}{ccccc}}     
\def\mxxxxxe{\end{array} \right)}
\def\kakkob{\left\{ \begin{array}{c}}
\def\kakkoe{\end{array}
            \right. }
\def\vecb{\left( \begin{array}{c}}
\def\vece{\end{array} \right) }
\def\vectttb{\left( \begin{array}{c}}
\def\vecttte{\end{array} \right) }
%
\def\dotf{\dot{\f}}
\def\dotr{\dot{\r}}
\def\dotff{\dot{f}}
\def\ddotf{\ddot{\f}}
\def\prif{\f '}
\def\prir{\r '}
\def\delp{\del _+}
\def\delm{\del _-}

\pagestyle{empty}
\setcounter{page}{1}

\Mfr{hep-th/9304145}

\Mfr{TEP-9R}

\Mfr{April 1993}\vss

\begin{center}

{\Huge A Bilocal Field Theory in }\vs{.7cm}

{\Huge Four Dimensions} \vs{4cm}

{\large {\bf Takayuki Hori }} \vs{0.7cm}

\begin{large}

{\it Institute of Physics, Teikyo University }  \vs{.4cm}

{\it Otsuka 359, Hachioji-shi, Tokyo 192-03, Japan } \vs{3.0cm}

{\bf Abstract}

\end{large}
\end{center}

\begin{normalsize}
\baselineskip=25pt

A bilocal field theory having M\"{o}bius gauge invariance is proposed.
In four dimensions there exists a zero momentum state of the first quantized
model, which belongs to a non-trivial BRS cohomology class.  A field theory
lagrangian having a gauge invariance only in four dimensions is constructed.
 \vs{.8cm}


\newpage

\def\ep{\varepsilon}
\def\d{\partial}

\pagestyle{plain}

The string model has achieved some impressive successes in unifying the
fundamental forces of nature. However, the proliferation of vacuum states
\cite{nara} makes it difficult to build the unique theory of everything.
Unfortunately, the structure of the model, in particular the superstring,
seems too complicated to construct a covariant non-perturbative framework
which may remove the degeneracy.

Nonetheless, the string model has brought us many remarkable properties,
{\it e.g.}, the critical dimension, the duality, the finiteness,
$E_{8}\times E_{8}$ etc. Among them the first two are the oldest and the
most characteristic features distinguishing the string model from others.
Does a truncated model of string exist, which possesses the above features?
The duality may be a consequence of $SL(2,R)$ (M\"{o}bius) symmetry and
possibly not of the full Virasoro algebra. It is conceivable that a system
with finite degrees of freedom, which has infinite components after the
second quantization, has the above properties. The rigid open string model
with the end points as the dynamical variables \cite{casa}, however, has
the larger
constraint algebra than $SL(2,R)$, and not a subalgebra of the Virasoro
algebra.

In this note we propose a bilocal field theory based on the bilocal particle
model \cite{tep8} whose constraint algebra is precisely $SL(2,R)$.
It is shown that the model has special properties in four space time
dimensions.
First there exists a zero momentum state in the first quantized model, which
is physical in four dimensions.
In the field theory a star product of two bilocal fields can be defined,
admitting the derivative law for the BRS operator and ensuring the gauge
invariance in four dimensions.
Thus an interacting field theory is constructed in an analogous way to
Witten's method in the covariant string field theory \cite{witt}.
Although we cannot claim the critical dimension of the model is four, we
get a general impression that the dimension four is special in the model.

Let us start with the following action of two relativistic particles in $D$
dimensions with mixing terms \cite{tep8}:
$$        I=\int d\tau [\frac{1}{2g_1}\dot{x}^{2}_{1}
                     +  \frac{1}{2g_2}\dot{x}^{2}_{2}
                     + e(\dot{x_{1}}x_{2} - x_{1}\dot{x}_{2} ) ],  \eqno(1)
$$
where $x_{a},g_a (a=1,2)$ are the coordinates of the particles and einbeins
of each world line, respectively, and the dots denote the derivatives with
respect to the world line parameter $\tau $ (we omit the space time indices
unless necessary). $e$ is the coupling constant with dimension of mass
square.  In the previous note [3] we showed that the action has a hidden
symmetry generated by
$$     \delta x_a = \ep _{a}\dot{x}_{a}
                 + \ep _{0}\sum_{b=1,2}s^{ab}\frac{\dot{x}_{b}}{g_b},
                                                                \eqno(2a)
$$
$$     \delta g_a = \frac{d}{d\tau }(\ep _{a}g_{a})
                    - (-1)^{a}4e\ep _{0}g_{a},
                                                                \eqno(2b)
$$
where $s^{12}=s^{21}=1, s^{11}=s^{22}=0$, and the local parameters $\ep $'s
are not independent but related by
$$         \dot{\ep }_{0} + 2eg_{1}g_{2}(\ep _{2} - \ep _{1}) = 0.
                                                                \eqno(3)
$$
When $e\neq 0$, $\ep_2$ is expressible in terms of $\ep_{0}$ and $\ep_{1}$
by Eq.(3), and the above transformations contain $\ep _0$ (and $\ep _1$)
and its derivatives with respect to $\tau $ up to second (first) orders.
Consequently, the physical degrees of freedom reduces by five. Two of them
fix $g_a$, leaving $2D-3$ components of coordinates being physical.  Direct
calculations show that the three degrees of gauge freedom relevant to the
unphysical coordinates form $SL(2,R)$ algebra.  The above facts are further
confirmed by the canonical formalism given below. If $e=0$, {\it i.e.},
the two relativistic particles are independent to each other, then the
physical degrees of freedom, of course, is $2D-2$. In this sense let us
call our system as a {\it bilocal particle} rather than two particles.

The canonical hamiltonian is given by
$$       {\cal H} = \lambda _{1}L_{1} + \lambda _{-1}L_{-1}
                    + \sum_{a=1,2}\Lambda _{a}\Pi _{a},
                                                             \eqno(4)
$$
where
$$       L_{1}  =  \frac{1}{4e}(p_{1} - ex_{2})^2,
                                                             \eqno(5a)
$$
$$       L_{-1} = -\frac{1}{4e}(p_{2} + ex_{1})^2.
                                                             \eqno(5b)
$$
$p_{a}$ and $\Pi _{a}$ are the canonical conjugates of $x_{a}$ and $g_{a}$,
respectively, and $\lambda $'s and $\Lambda $'s are the lagrange
multipliers.
The canonical equations coincide with the Euler equations if one puts
$\lambda _{1}=2eg_1$, $\lambda _{-1}=-2eg_2$ and $\Lambda _{a}=\dot{g_a}$.
But they are arbitrary functions of canonical variables in the Dirac
formalism \cite{dira}, which merely fix the gauge. Hereafter we choose the
gauge $\Lambda _{a}=0$ corresponding to $\dot{g_a}=0$. Preservation of the
primary constraint, $\Pi _{a}=0$, along the $\tau $ development requires the
secondary constraint, $L_{\pm 1}=0$ whose preservation, in turn, requires
the tertiary constraint,
$$       L_{0} = -\frac{i}{4e}(p_{1} -ex_{2})(p_{2} + ex_{1}) = 0.
                                                               \eqno(6)
$$
Finally $\dot{L_0}=0$ is guaranteed by $L_{\pm 1}=0$. Corresponding to the
constraints, $L_{0,\pm 1}=\Pi _{1,2}=0$, we have five unphysical canonical
pairs, and the number of physical coordinates is $2D-3$ as is declared above.

Now let us proceed to the quantum theory. Replacing $p_{a}$ by
$-i\frac{\d }{\d x_a}$ in $L_a$ we get the quantum operators $\hat{L}_{0}$
and $\hat{L}_{\pm 1}$ which form $SL(2,R)$:
$$  [\hat{L}_{n},\hat{L}_{m}] = (n-m)\hat{L}_{n+m} +
          2(a-\frac{D}{4})\delta _{n+m},
          (n,m=0,\pm 1),
                                                              \eqno(7)
$$
where the central term is merely attributed to the ordering ambiguity of
$\hat{L}_{0}$ which we write as
$\hat{L}_{0} = -\frac{i}{4e}(-i\d _{1} - ex_{2})(-i\d _{2} + ex_{1}) - a$.
The constant $a$ plays no role because of the finite degrees of freedom,
henceforth we choose $a=\frac{D}{4}$ . The nilpotent BRS operator is
written in this notation of $\hat{L}_{0}$ as
$$      Q = \sum_{n=0,\pm 1}c_{n}\hat{L}_{n} +
              \frac{i}{2}\sum_{n,m=0,\pm 1}(n-m)c_{n}c_{m}{\cal P}_{n+m},
                                                                \eqno(8)
$$
where $c$'s are ghost variables and ${\cal P}_{n}=i\frac{\d }{\d c_n}$. The
requirement of the nilpotency of $Q$ removes the ordering ambiguity in
$Q$ at all, making $Q$ independent on any arbitrary parameters. (If we
choose, say, $a=137$, then we have a central term in the algebra defined
in Eq.(7), and by the requirement of the nilpotency the BRS operator
acquires the extra term, $(137-\frac{D}{4})c_{0}$, compensating the shift
in $\hat{L}_{0}$. Of course, {\it e.g.}, Eq.(17) derived later is
unchanged.)
This is the common property of any systems with finite degrees of freedom.
Note the space time dimension is not specified by the nilpotency as opposed
to the string model \cite{kato}.

The physical states are defined by the Kugo-Ojima condition \cite{kugo},
$$       Q\mid phys\rangle  = 0.                              \eqno(9)
$$
At first sight the physical state condition would amount to vanishing all
$\hat{L}$'s as they operate to $\mid phys\rangle $, since in our notation
the central term in Eq.(7) is absent.
However, there exists an alternative scheme which is possible in four
dimensions as is described later.
described la
This is caused by an unusual kinematical symmetry as is shown below.

The space time translations and the Lorentz transformations are generated by
$$          \tilde{p}_{+} = p_1 + ex_2,                     \eqno(10a)
$$
$$          \tilde{p}_{-} = p_2 - ex_1,                     \eqno(10b)
$$
$$          M_{\mu \nu} = \sum_{a=1,2}p_{a[\mu}x_{a\nu ]},  \eqno(10c)
$$
where $p_{a}=-i\d _{a}$. The validity of Eqs.(10) is confirmed by the
defying equations of the momenta in terms of $x$'s and $\dot{x}$'s. It is
important to note the difference of signs between Eqs.(10a,b) and
Eqs.(5a,b). Thus the motions of the two particles are not necessarily
light-like by the constraints, $\hat{L}_{0,\pm 1}=0$. The generators
satisfy the algebra
$$          [M_{\mu \nu },M_{\lambda \rho }] =
i\eta _{\rho [\mu}M_{\nu ]\lambda} - i\eta _{\lambda [\mu}M_{\nu ]\rho},
                                                          \eqno(11a)
$$
$$          [M_{\mu \nu}, \tilde{p}_{\pm \lambda}] =
                     -i\eta _{\lambda [\mu}
                    \tilde{p}_{\pm \nu ]},
                                             \eqno(11b)
$$
$$          [\tilde{p}_{+\mu},\tilde{p}_{-\nu}] = 2ei\eta _{\mu \nu}.
                                                      \eqno(11c)
$$
It is a quite important fact that the momenta of each particle,
$\tilde{p}_{\pm}$, cannot have definite values simultaneously because
of the {\it uncertainty relation} (11c). The physical states should
belong to irreducible representations of the algebra, denote ${\cal P}$,
defined by Eqs.(11), which contains the Poincar\'{e} algebra.

A short calculation shows that
$$          [\tilde{p}_{\pm \mu }, \hat{L}_{0,\pm 1}] =
                     [M_{\mu \nu}, \hat{L}_{0,\pm 1}] = 0.
                                              \eqno(12)
$$
This means that the translations and the Lorentz transformations of the
hamiltonian ${\cal H}$ amount to changes of the Lagrange multipliers,
{\it i.e.}, the gauge choice. A state belonging to an irreducible
representation of ${\cal P}$ is transformed by the operation of
$\hat{L}_{0,\pm 1}$ to a gauge equivalent state. Since the $SL(2,R)$
commutes with ${\cal P}$ the gauge equivalent states have common
kinematical quantum numbers.

Now the ground state is defined by
$$      M_{\mu \nu}\mid 0\rangle  = \tilde{p}_{\mu}\mid 0\rangle  = 0,
                                                           \eqno(13)
$$
where $\tilde{p}=\tilde{p}_{+}+\tilde{p}_{-}$ is the total momentum.
We call the ground state as {\it vacuum}, since it has the quantum number
of the Poincar\'{e} vacuum.
Let us consider the representations on the configuration space spanned by
($x^{\mu}_{a},c_{0,\pm 1}$). By solving Eqs.(13) we obtain
$$            \mid 0\rangle  = Fe^{\frac{ie}{2}(x^{2}_{1} - x^{2}_{2})},
                                                               \eqno(14)
$$
where $F$ is an arbitrary function of $(x_{1} - x_{2})^{2}$ and the ghost
variables. Note that the translational invariant state is not a constant
but expressed by the non trivial function of $x_{a}$. This is a consequence
of the fact that the total momentum is the sum of the mutually
non-commuting momenta of two particles.

The vacuum state should not necessarily be a physical state as in the case
of the string model where the ground state of the first quantized model is
generally unphysical.
In fact the vacuum is not physical if it has zero ghost number as we see
shortly.
On the other hand, if the vacuum is physical we can construct a physical
{\it Fock space} by acting kinematic operators on the vacuum as is shown
later.
So let us seek the possibility of existence of the physical vacuum with
non-zero ghost number.

We assume that the vacuum is a non-trivial element of the BRS cohomology,
{\it i.e.}, it satisfies the physical state condition {\it and} is not
written as $Q\mid \chi \rangle $ for some $\mid \chi \rangle $.
There are four possibilities according to the ghost number of the vacuum,
defined by $N_{g}=-i\sum_{n=0,\pm 1}c_{n}{\cal P}_n$. If $N_{g}(vac)=0$,
then all $\hat{L}_{a} (a=0,\pm 1)$ must vanish as operating on the vacuum.
However, we see this is impossible as follows. $F$ is determined only by
$\hat{L}_{1}\mid 0\rangle =0$, and the vacuum would be expressed as
$$             \mid 0 -{\rangle }_{0} = e^{ie(x_{1} - x_{2})x_{2}}.
                                                             \eqno(15a)
$$
But $\hat{L}_{0,-1}\mid 0\rangle =0$ cannot be satisfied any more. For
later convenience we write also the solution of
$\hat{L}_{-1}\mid 0\rangle =0$:
$$             \mid 0 +{\rangle }_{0} = e^{ie(x_{1} - x_{2})x_{1}},
                                                           \eqno(15b)
$$
which does not vanish by the operations of $\hat{L}_{1}$ and
$\hat{L}_{0}$.  Next possibility is $N_{g}(vac)=1$. Write
$\mid 0\rangle $ as $\sum_{n=0,\pm 1}c_{n}A_{n}$. If $A_{0}\ne 0$,
then the physical state condition is solved as
$\mid 0\rangle =Q\hat{L}_{0}^{-1}A_{0}$, {\it i.e.}, a trivial element
of the BRS cohomology. (In showing this we have used the relation,
$\hat{L}_{\pm 1}f(\hat{L}_{0})=f(\hat{L}_{0}\pm 1)\hat{L}_{\pm 1}$ for
any function $f$.) If $A_{0}=0$, then it follows that
$\hat{L}_{-1}A_{1}=\hat{L}_{1}A_{-1}=0$. Then, the vacuum must be
written as a linear combination of
$$               \mid 0 -{\rangle }_{1} = c_{-1}\mid 0 -{\rangle}_{0},
                                                            \eqno(16a)
$$
and
$$               \mid 0 +{\rangle}_{1} = c_{1}\mid 0 +{\rangle}_{0}.
                                                             \eqno(16b)
$$
The physical state condition, however, is not satisfied automatically. In
fact we see
$$            Q\mid 0 \pm {\rangle}_{1} =
               \mp (\frac{D}{4} - 1)c_{0}\mid 0 \pm {\rangle}_{1}.
                                                            \eqno(17)
$$
Hence, if the dimension of the space time is four we have the non-trivial
solutions with ghost number one, expressed in Eqs.(16a) and (16b).
(Needless to say that Eq.(17) is not a consequence of the particular
choice $a=\frac{D}{4}$.)  Next we assume $N_{g}(vac)=2$. The vacuum must
be of the form:
$$               \mid 0 \pm{\rangle}_{2} = c_{0}\mid 0 \pm{\rangle}_{1}.
                                                             \eqno(18)
$$
These states satisfy the physical state condition in arbitrary dimensions.
If $D\ne 4$, however, these states are trivial elements of the BRS
cohomology as is seen from Eq.(17). Since there are three ghost variables
the final possibility is $N_{g}(vac)=3$. Any states with ghost number three
satisfy the physical state condition. But it is easy to see that they are
all trivial elements of the BRS cohomology.
Therefore, we conclude that the assumption of the physical state condition
for the vacuum specifies the dimension of the space time to be four, the
physical dimension!
It must be stressed, however, that this does not mean the model is proved to
have critical dimension, since the ordinary unphysical vacuum exists in any
dimensions.

\def\pt{\tilde{p}}
\def\Lm{\hat{L}_{-1}}
\def\Lp{\hat{L}_{1}}
\def\Lz{\hat{L}_0}
\def\Lpm{\hat{L}_{\pm 1}}

Now let us give some speculations on the construction of a field theory.
We want to obtain a functional action in terms of bilocal fields,
{\it i.e.}, arbitrary functions of $x_1, x_2$ and the ghost variables.
It is convenient to expand a bilocal field in terms of a complete system
of base functions.
As is well known in the compact group one can construct a system of base
functions in a function space, the completeness of which in $L_2$, the set
of all square integrable functions, is guaranteed by the Peter-Wyle theorem
on the compact topological group.
In our case it is natural to employ the algebra, ${\cal P}$, defined in
Eq.(11), for the purpose of constructing the complete system, since
${\cal P}$ generates the global symmetry of the first-quantized action and
should provide a base for defining ``kinematic'' quantum numbers.

Before defining the field theory let us consider the representation of
${\cal P}$, which is constructed on specific vacuums.
In four dimensions the vacuum has the four fold degeneracy expressed in
Eqs.(16a,b) and (18).
We discard, for the moment, the vacuum states with ghost number two and the
one having $c_1$ as the ghost factor.
The only quantities in our hands, which would create new states by
operating on the vacuum, are $x_1, x_2, p_1$ and $p_2$.
Instead of these sixteen quantities we can use the fourteen  generators of
${\cal P}$ and the two $\Lpm$ by redefining the independent variables
(note that this is possible only in four dimensions).
Then it turns out that the general states obtained by operating the above
quantities repeatedly on the vacuum, $\mid 0 -\ket$, are linear
combinations (or infinite sums) of the following states:
$$           \mid n {\ket}^{\mu _1...\mu _i} =
                \Lm ^n\pt ^{\mu _1}_+...\pt ^{\mu _i}_+\mid 0 -{\ket}_1 .
                                                               \eqno(19)
$$
That is, $\pt ^{\m}_-$ and $M^{\m \n}$ are redundant operators.
This is verified as follows.
First note that the scalar state with total momentum $k$ is
$$           \mid k\ket  = e^{ikx_2}\mid 0 -{\ket}_{1} =
               e^{\fr{i}{2e}k\pt _{+}}\mid 0 -{\ket}_{1}.
                                                               \eqno(20)
$$
Then, by using Eq.(11c), we see the state
$\pt ^{\m}_-\mid 0 -{\ket}_1 = \pt ^{\m}_-e^{-\fr{i}{2e}k\pt _+}\mid k\ket$
is a linear combination (actually an infinite sum) of the states of the
form $\pt ^{\mu _1}_+...\pt ^{\mu _i}_+\pt ^{\mu}_-\mid k\ket$.
The latter states are further expressed as linear combinations of the
states defined by Eq.(19) with
$n = 0$, since $\pt ^{\m}_-\mid k\ket = (k^{\m} - \pt ^{\m}_+)\mid k\ket$.
By using Eq.(11c) again, we see the operator $\pt ^{\m}_-$ can be always
eliminated if it would appear in the definition, Eq.(19), of the general
states.
Similarly, by using $M^{\m \n}\mid k\ket  \sim  \mid k'\ket$, we see
$M^{\m \n}$ are also redundant operators.
Since $\{ \Lz , \Lpm \}$ commute with the generators of ${\cal P}$ and
$\Lp \mid 0 -{\ket}_1 = 0$ the most general state constructed from
$\mid 0 -\ket$ is given by Eq.(19), which completes the proof.

The vector space spanned by the system defined by Eq.(19) forms a
(reducible) representation of ${\cal P}$.
Replacing the minus (plus) signs in the above equations by the plus
(minus) signs we obtain another representation, and the direct sum of the
two representations should exhaust all possible states constructed from
the two vacuums $\mid 0 \pm{\ket}_1$.
These representations are divided into invariant representations by
specific values of $n$, since $\{ \Lz , \Lpm \}$ commute with the
generators of ${\cal P}$.
The representation with $n = 0$ consists of the physical states, and ones
with $n \ne  0$ consist of the unphysical states.

On the base of the above argument we conjecture a theorem like the
Peter-Wyle, which would claim the completeness of the system defined by
$$           u_{n}^{\pm  \mu _1...\mu _i} =
    \Lpm ^n\pt ^{\mu _1}_{\mp}...\pt ^{\mu _i}_{\mp}\mid 0 \pm {\ket}_0 .
                                                               \eqno(21)
$$
in a sufficiently large space of functions on $(x_1, x_2)$, where we
treat the ghost variables separately.
The basis (21) also forms a representation of $SL(2, R)$ generated by
$\{ \Lz , \Lpm \}$.
In particular we have
$$          \Lz u_{n}^{\pm \m _1...\m _k} =
       \mp (n + \fr{D}{4})u_{n}^{\pm \m _1...\m _k},
                                                          \eqno(22a)
$$
$$          \Lpm u_{0}^{\mp \m _1...\m _k} = 0,
                                                          \eqno(22b)
$$
where  we keep $D$ in the expressions here and hereafter in order to
insist that  the value $D = 4$ is special also in the field theory.

Leaving aside mathematical problems like the completeness of the system
and the complete classification of the irreducible representations of
${\cal P}$ for a future study, let us proceed tentatively to construction
of a field theory.
We assume that a bilocal field is expanded as
$$    A(x_{1},x_{2},c) =
\sum_{a=\pm }\sum_{n=0}^{\infty}\sum_{\{ \m _i\}
}\fr{1}{n!}A_{n\m _1...\m _k}^{a}(c)u_{n}^{a\m _1...\m _k}(x_{1},x_{2}).
                                                             \eqno(23)
$$
In order to obtain an action for bilocal fields in the analogous manner as
Witten's string field theory \cite{witt} let us define the star product of
two bilocal fields as follows:
$$         A*B = \sum_{a=\pm }\sum_{n=0}^{\infty }\sum_{m=0}^{n}\sum_{k =
0}^{\infty}\sum_{i = 0}^{k}\frac{1}{n!}A_{n-m, \m _1...\m
_i}^{a}B_{m-\frac{D}{4}, \m _{i + 1}...\m _{k}}^{a}u_{n}^{a, \m
_1...\m _k},
                                                             \eqno(24)
$$
where $A_n = B_n = 0$ for $n < 0$.
The factor $\frac{1}{n!}$, in Eqs.(23) and (24), plays some role in what
follows.  Note the definition of the star product is meaningful only when
$\frac{D}{4}$ is integer.  The star product is associative and commutative
up to the Grassmann parity.  Using some relations like $[\hat{L}_{\pm
1},\hat{L}_{\mp 1}^{n}]=n\hat{L}_{\mp 1}^{n-1}(n-1\pm 2\hat{L}_{0})$ we can
show the {\it Leibniz rule} for $Q$:
\setcounter{equation}{24}
\bena
  & &     Q(A*B) = QA*B + (-1)^{N_{A}}A*QB \nn
  &+& (\frac{D}{4} - 1)\sum_{a=\pm }\sum_{n=0}^{\infty
}\sum_{m=0}^{n}\sum_{k = 0}^{\infty}\sum_{i =
0}^{k}\frac{(a)c_a}{n!}A_{n-m+1, \m _1...\m _i}^{a}B_{m-\frac{D}{4},
\m _{i + 1}...\m _{k}}^{a}u_{n}^{a, \m _1...\m _k},
\eena
where $N_{A}$ is the Grassmann parity of $A$.  Surprisingly $Q$ behaves
as a derivative operator only if $D=4$.

We assume the bilocal field is Grassmann odd and has the Yang-Mills
indices.  Equipped with the above tools we can easily write the action
of the bilocal field theory as
$$    I = \int d^{4}x_{1}d^{4}x_{2}dc_{1}dc_{-1}dc_{0}V
          [{\cal A}_{i}*Q{\cal A}_{i}  - \frac{1}{3}gf^{ijk}{\cal
A}_{i}*{\cal A}_{j}*{\cal A}_{k}],
                                                            \eqno(26)
$$
where $f^{ijk}$ is the structure constants (with indices raised by the
Killing form) and $g$ is the coupling constant. $V=V(x_{1},x_{2})$ is a
measure factor satisfying $\tilde{p}^{2}_{+}V=\tilde{p}^{2}_{-}V=0$,
which guarantees the integral of the {\it total derivative}, $QA$, to
vanish. Employing the derivative law for $Q$, which holds only when
$D=4$, the action is shown to be invariant under the gauge
transformations defined by
$$  \delta {\cal A}_{i}  = Q{\L}_{i}  + gf_{i}^{jk}{\L}_{j}*{\cal A}_{k},
                                                            \eqno(27)
$$
where $\L$'s are arbitrary functions of $(x_{1,2},c_{0,\pm 1})$,
which are Grassmann even. ${\cal A}$ contains the sectors with various
ghost numbers up to three.  Among them the sector with ghost number
one is of special interest, since it contains the gauge fields of
ordinary formulation as is seen from Eq.(27). This sector contains the
non-gauge physical fields as well, and the kinetic terms are closed in
itself. The sectors with ghost number zero or two do not contain gauge
fields, but propagate since they have mixed kinetic terms in Eq.(26).
They consist of fermions with integer spins. The fields with ghost
number three are all auxiliary ones, since they have not kinetic terms.

In view of the above remarks we see that the simplest framework is
obtained by restricting ourselves to the sector with ghost number one.
In this sector the Hilbert space determined by the asymptotic fields is
constructed on the vacuum states $\mid 0 \pm {\rangle}_{1}$ appeared in
the first quantized theory outlined before.  \vspace{0.7cm}

I am grateful to S. Saito and M. Kamata for stimulating discussions.
\newpage


\newpage
\end{normalsize}
\end{document}